\documentclass[aps,preprint]{revtex4}
\usepackage{amssymb,epsf}
\usepackage{amsmath}
\usepackage{graphicx}

\begin{document}
\title{Thermodynamics of rotating black branes in $(n+1)$-dimensional
Einstein-Born-Infeld gravity}
\author{M. H. Dehghani$^{1,2}$\footnote{email address:
mhd@shirazu.ac.ir} and H. R. Rastegar Sedehi$^{1}$}
\affiliation{$^1$Physics Department and Biruni Observatory,
College of Sciences, Shiraz
University, Shiraz 71454, Iran\\
$^2$Research Institute for Astrophysics and Astronomy of Maragha
(RIAAM), Maragha, Iran}

\begin{abstract}
We construct a new class of charged rotating solutions of
$(n+1)$-dimensional Einstein-Born-Infeld gravity with cylindrical
or toroidal horizons in the presence of cosmological constant and
investigate their properties. These solutions are asymptotically
(anti)-de Sitter and reduce to the solutions of Einstein-Maxwell
gravity as the Born-Infeld parameters goes to infinity. We find
that these solutions can represent black branes, with inner and
outer event horizons, an extreme black brane or a naked
singularity provided the parameters of the solutions are chosen
suitably. We compute temperature, mass, angular momentum, entropy,
charge and electric potential of the black brane solutions. We
obtain a Smarr-type formula and show that these quantities satisfy
the first law of thermodynamics. We also perform a stability
analysis by computing the heat capacity and the determinant of
Hessian matrix of mass with respect to its thermodynamic variables
in both the canonical and the grand-canonical ensembles, and show
that the system is thermally stable in the whole phase space.
\end{abstract}
 \maketitle

\section{Introduction}

In recent years, a great deal of attention has been focused on the
thermodynamics of asymptotically anti-de Sitter (AAdS) black
holes. One reason for this is the role of AdS/CFT duality
\cite{Mal}, and in particular, Witten interpretation of the
Hawking-Page phase transition between thermal AdS and
asymptotically AdS black hole as the confinement-deconfinment
phases of the Yang-Mills theory defined on the asymptotic
boundaries of the AdS geometry \cite{Wit}. The thermodynamics of
black holes may be modified in the presence of matter fields. For
example a scalar field makes the $(n+1)$-dimensional rotating
solutions of Einstein gravity unstable \cite{Deh1}, while the
presence of electromagnetic field \cite{Deh2} or higher curvature
terms \cite{Deh3} have not any effects on the stability of the
solutions. The fact that thermodynamics of black holes may be
modified in the presence of matter fields provides a strong
motivation for considering thermodynamics of black holes in the
presence of non-linear electromagnetic field. In this paper we
consider the thermodynamics of rotating black branes in the
presence of non-linear electromagnetic field. The pioneering
theory of the non-linear electromagnetic field was formulated by
Born and Infeld (BI) in 1934 \cite {BI}. Their basic motivation
was to solve the problem of the self-energy of the electron by
imposing a maximum strength for the electromagnetic field.
Although their attempt did not succeed in this regard, many kinds
of solution, such as a vortex and a particle-like solution, which
were constructed afterward in the model including BI term were of
great interest. In 1935, Hoffmann \cite{Hof} joined general
relativity with Born-Infeld electrodynamics to obtain a
spherically symmetric solution representing the gravitational
field of a charged object. These works were nearly forgotten for
several decades, until the interest in non-linear electrodynamics
increased in the context of low energy string theory, in which
Born-Infeld type actions appeared \cite{Frod}.

The spherically symmetric solutions in Einstein-Born-Infeld (EBI)
gravity with or without a cosmological constant have been
considered by many authors \cite{EBI,Cai,Dey}. In this paper we
are interested in the thermodynamics of asymptotically anti-de
Sitter (AAdS) rotating solutions of EBI gravity. The AAdS rotating
solution of Einstein's equation with cylindrical and toroidal
horizon and its extension to include the linear electromagnetic
field have been considered in Ref. \cite{Lemos}. The
generalization of this AAdS charged rotating solution of
Einstein-Maxwell gravity to the higher dimensions has been done in
Ref. \cite{Awad}. Thermodynamics of these solutions have been
investigated in Ref. \cite{Deh2}. Also, these kinds of charged
rotating solutions have been introduced in Lovelock gravity and
their thermodynamics have been investigated \cite{Deh3}.

The outline of our paper is as follows. We give a brief review of
the field equations of EBI gravity and present exact rotating
solutions of these field equations in Sec. \ref{Sol}. We also
investigate their properties and obtain their thermodynamic and
conserved quantities through the use of the counterterm method. In
Sec. \ref{Therm}, we obtain a Smarr-type formula and show that
these quantities satisfy the first law of thermodynamics. We also
perform a local stability analysis of the black holes in the
canonical and grand canonical ensembles. We finish our paper with
some concluding remarks.

\section{Rotating Black Brane in EBI Gravity\label{Sol}}

The gravitational action for Einstein gravity in the presence of
Born-Infeld field with a cosmological constant $\Lambda $ is
\begin{equation}
I_{G}=-\frac{1}{16\pi }\int d^{n+1}x\sqrt{-g}\left( {\cal
R}-2\Lambda
+L(F)\right) +\frac{1}{8\pi }\int_{\partial {\cal M}}d^{n}x\sqrt{_{-}\gamma }%
K(\gamma ),  \label{Actg}
\end{equation}
where $L(F)$ is the Lagrangian of BI field given as
\begin{equation}
L(F)=4\beta ^{2}\left( 1-\sqrt{1+\frac{F^{2}}{2\beta ^{2}}}\right)
. \label{LagBI}
\end{equation}
In Eq. (\ref{LagBI}) $F^{2}=F^{\mu \nu }F_{\mu \nu }$, where
$F_{\mu \nu }=\partial _{\mu }A_{\nu }-\partial _{\nu }A_{\mu }$
is the electromagnetic tensor field, $A_{\mu }$ is the vector
potential, and $\beta $ is the Born-Infeld parameter with
dimension of mass. In the limit $\beta
\rightarrow \infty $, $L(F)$ reduces to the standard Maxwell Lagrangian $%
L(F)=-F^{2}$, while $L(F)\rightarrow 0$ as $\beta \rightarrow 0$.
The first integral in Eq. (\ref{Actg}) is the Einstein-Hilbert
volume term with negative cosmological constant $\Lambda
=-n(n-1)/2l^{2}$ in the presence of BI field, and the second
integral is the Gibbons Hawking boundary term which
is chosen such that the variational principle is well-defined. The manifold $%
{\cal M}$ has metric $g_{\mu \nu }$ and covariant derivative $\nabla _{\mu }$%
. $K$ is the trace of the extrinsic curvature $K^{\mu \nu }$ of
any boundary(ies) $\partial {\cal M}$ of the manifold ${\cal M}$,
with induced metric(s) $\gamma _{ij}$.

Varying the action with respect to the gauge field $A_{\mu }$ and
the gravitational field $g_{\mu \nu }$, the field equations are
obtained as
\begin{equation}
\partial _{\mu }\left( \frac{\sqrt{-g}F^{\mu \nu }}{\sqrt{1+\frac{F^{2}}{%
2\beta ^{2}}}}\right) =0,  \label{BIeq}
\end{equation}
\begin{equation}
G_{\mu \nu }+\Lambda g_{\mu \nu }=\frac{1}{2}g_{\mu \nu
}L(F)+\frac{2F_{\mu \lambda }F_{\nu }^{~\lambda
}}{\sqrt{1+\frac{F^{2}}{2\beta ^{2}}}}, \label{EBIeq}
\end{equation}
where $G_{\mu \nu }$ is the Einstein tensor.

The metric of $(n+1)$-dimensional AAdS charged rotating black
brane with $k$ rotation parameters $a_{i}$'s is \cite{Awad}
\begin{eqnarray}
ds^{2} &=&-f(r)\left( \Xi dt-{{\sum_{i=1}^{k}}}a_{i}d\phi _{i}\right) ^{2}+%
\frac{r^{2}}{l^{4}}{{\sum_{i=1}^{k}}}\left( a_{i}dt-\Xi l^{2}d\phi
_{i}\right) ^{2}  \nonumber \\
&&\ \text{ }+\frac{dr^{2}}{f(r)}-\frac{r^{2}}{l^{2}}{\sum_{i<j}^{k}}%
(a_{i}d\phi _{j}-a_{j}d\phi _{i})^{2}+r^{2}dX^{2},  \label{met}
\end{eqnarray}
where $\Xi =\sqrt{1+\sum_{i}^{k}a_{i}^{2}/l^{2}}$ and $dX^{2}$ is
the Euclidean metric on the $\left( n-1-k\right) $-dimensional
submanifold with volume $\Sigma _{n-k-1}$. The maximum value of
$k$ is $[n/2]$, where $[x]$ denotes the integer part of $x$.
First, we use the gauge potential ansatz
\begin{equation}
A_{\mu }=h(r)\left( \Xi \delta _{\mu }^{0}-\delta _{\mu
}^{i}a_{i}\right) \text{(no sum on }i\text{)}  \label{Amu}
\end{equation}
in the non-linear electromagnetic field equation (\ref{BIeq}). We
obtain
\begin{equation}
h(r)=-\sqrt{\frac{n-1}{2n-4}}\frac{q}{r^{n-2}}\ {_{2}F_{1}\left( \left[\frac{%
1}{2},\frac{n-2}{2n-2}\right],\left[\frac{3n-4}{2n-2}\right],-\eta
\right)}, \label{hr}
\end{equation}
where $q$ is an integration constant which is related to the
charge parameter, $_{2}F_{1}([a,b],[c],z)$ is hypergeometric
function and
\[
\eta =\frac{{(n-1)(n-2)q^{2}}}{2\beta ^{2}r^{2n-2}}.
\]
One may note that ${_{2}F_{1}\rightarrow 1}$ as $\eta \rightarrow
0$ ($\beta \rightarrow \infty $) and $A_{\mu }$ of Eq. (\ref{Amu})
reduces to the gauge potential of Maxwell field \cite{Deh2}. To
find the function $f(r)$ , one
may use any components of Eq. (\ref{EBIeq}). The simplest equation is the $%
rr $ component of these equations which can be written as
\begin{equation}
(n-1)r^{n-2}f^{\prime }+(n-1)(n-2)r^{n-3}f+2\left[ \Lambda +2{\beta }%
^{2}\left( \left( 1+\eta \right) ^{1/2}-1\right) \right]
r^{n-1}=0, \label{rreq}
\end{equation}
where the prime denotes a derivative with respect to $r$. The
solutions of Eq. (\ref{rreq}) can be written as
\begin{eqnarray}
f(r) &=&\frac{r^{2}}{l^{2}}-\frac{m}{r^{n-2}}+{\frac{4{\beta }^{2}r^{2}}{%
n(n-1)}}\left[ 1-\left( 1+\eta \right) ^{1/2}\right]  \nonumber \\
&&+{\frac{2(n-1)q^{2}}{nr^{2n-4}}}\ _{2}F_{1}\left( \left[\frac{1}{2},\frac{%
n-2}{2n-2}\right],\left[\frac{3n-4}{2n-2}\right],-\eta \right)
\label{fr}
\end{eqnarray}
Although the other components of the field Eq. (\ref{EBIeq}) are
more complicated, one can check that the metric (\ref{met})
satisfies all the
components Eq. (\ref{EBIeq}) provided $f(r)$ is given by (\ref{fr}). Again, $%
f(r)$ reduces to the metric function of Ref. \cite{Deh2} as $\beta
$ goes to infinity.

\subsection{Properties of the solutions}

The solutions of Eqs. (\ref{met}) and (\ref{fr}) are
asymptotically AdS. One can show that the Kretschmann scalar
$R_{\mu \nu \lambda \kappa }R^{\mu \nu \lambda \kappa }$ diverges
at $r=0$, and therefore there is a curvature singularity located
at $r=0$. Seeking possible black hole solutions, we turn to look
for the existence of horizons. As in the case of rotating black
hole solutions of Einstein-Maxwell gravity, the above metric given
by Eq. (\ref {met}) has two types of Killing and event horizons.
The Killing horizon is a null surface whose null generators are
tangent to a Killing field. It was proved that a stationary black
hole event horizon should be a Killing horizon in the
four-dimensional Einstein gravity \cite{Haw1}. This fact is also
true for this $(n+1)$-dimensional metric and the Killing vector
\begin{equation}
\chi =\partial _{t}+{\sum_{i}^{k}}\Omega _{i}\partial _{\phi
_{i}}, \label{Kil}
\end{equation}
is the null generator of the event horizon.

One can obtain the temperature and angular momentum of the event
horizon by
analytic continuation of the metric. Setting $t\rightarrow i\tau $ and $%
a_{i}\rightarrow ia_{i}$ yields the Euclidean section of
(\ref{met}), whose regularity at $r=r_{+}$ requires that we should
identify $\tau \sim \tau
+\beta _{+}$ and $\phi _{i}\sim \phi _{i}+\beta _{+}\Omega _{i}$, where $%
\beta _{+}$ and $\Omega _{i}$'s are the inverse Hawking
temperature and the angular velocities of the outer event horizon.
One obtains
\begin{eqnarray}
T_{+} &=&\frac{f^{\prime }(r_{+})}{4\pi \Xi }={{\frac{nr_{+}}{4\pi l^{2}\Xi }%
}}\left( {1+{\frac{4\beta ^{2}l^{2}}{n(n-1)}}\left[ 1-\left(
1+\eta_+ \right)
^{1/2}\right] }\right) ,  \label{Temp} \\
\Omega _{i} &=&\frac{a_{i}}{\Xi l^{2}}.  \label{Om}
\end{eqnarray}
Using the fact that the temperature of extreme black hole is zero,
one can obtain $r_{+}$ in term of the charge parameter $q_{{\rm
ext}}$ as
\begin{equation}
r_{+}=\left( \frac{8(n-2)\beta ^{2}l^{4}q_{{\rm ext}}^{2}}{n^{2}(n-1)+8n{%
\beta }^{2}l^{2}}\right) ^{\frac{1}{2n-2}}  \label{qext}
\end{equation}
Inserting the above $r_{+}$ in the equation $f(r_{+})=0$, the
condition for having an extreme black brane is obtained as
\begin{eqnarray}
m_{{\rm ext}} &=&q_{{\rm ext}}^{n/(n-1)}\Big[\left(
\frac{(n-1)(n-2)}{2\beta
^{2}\eta _{{\rm ext}}}\right) ^{n/(2n-2)}\left( \frac{1}{l^{2}}+{\frac{4{%
\beta }^{2}}{n(n-1)}}\left( 1-\left( 1+\eta _{{\rm ext}}\right)
^{1/2}\right) \right)+  \nonumber \\
&&{\frac{2n-2}{n}\left( \frac{(n-1)(n-2)}{2\beta ^{2}\eta _{{\rm ext}}}%
\right)^{-(n-2)/(2n-2)}}\,_{2}F_{1}\left( \left[\frac{1}{2},\frac{n-2}{2n-2}%
\right],\left[\frac{3n-4}{2n-2}\right],-\eta _{{\rm ext}}\right)\Big]; \nonumber \\
\eta _{{\rm ext}} &=&{\frac{n(n-1)\left[ n(n-1)+8{\beta }^{2}l^{2}\right] }{%
16\beta ^{4}l^{4}}}
\end{eqnarray}
The metric of Eqs. (\ref{met}) and (\ref{fr}) presents a black
brane solution with inner and outer horizons, provided the mass
parameter $m$ is greater than $m_{{\rm ext}}$, an extreme black
hole for $m=m_{{\rm ext}}$ and a naked singularity otherwise.

Usually the entropy of the black holes satisfies the so-called
area law of entropy in Einstein gravity which states that the
black hole entropy equals to one-quarter of horizon area
\cite{Beck}. It applies to almost all kind of black holes and
black branes \cite{Haw2}. Denoting the volume of the hypersurface
boundary at constant $t$ and $r$ by $V_{n-1}=(2\pi )^{k}\Sigma
_{n-k-1}$, it is a matter of calculation to show that the entropy
per unit volume $V_{n-1}$ of the black brane is
\begin{equation}
S=\frac{\Xi }{4}r_{+}^{(n-1)}  \label{Entropy}
\end{equation}
The charge of the black brane per unit volume $V_{n-1}$ can be
found by calculating the flux of the electric field at infinity,
yielding
\begin{equation}
Q=\frac{\Xi }{4\pi }\sqrt{\frac{(n-1)(n-2)}{2}}q.  \label{Charg}
\end{equation}
The electric potential $\Phi $, measured at infinity with respect
to the horizon, is defined by \cite{Gub}
\begin{equation}
\Phi =A_{\mu }\chi ^{\mu }\left| _{r\rightarrow \infty }-A_{\mu
}\chi ^{\mu }\right| _{r=r_{+}},  \label{Pot1}
\end{equation}
where $\chi $ is the null generator of the horizon given by Eq.
(\ref{Kil}). One finds
\begin{equation}
\Phi =\sqrt{\frac{(n-1)}{2(n-2)}}\frac{q}{\Xi {r_{+}}^{n-2}}\ {%
_{2}F_{1}\left( \left[\frac{1}{2},\frac{n-2}{2n-2}\right],\left[\frac{3n-4}{%
2n-2}\right],-\eta_{+} \right)}.  \label{Pot}
\end{equation}

The conserved charges of the black brane may be found through the
use of the counterterm method inspired by the AdS/CFT
correspondence. The finite stress-energy tensor in
$(n+1)$-dimensional Einstein gravity with flat boundary is
\begin{equation}
T^{ab}=\frac{1}{8\pi }\left[ \Theta ^{ab}-\Theta
h^{ab}+\frac{n-1}{l}\gamma ^{ab}\right] ,  \label{Stres}
\end{equation}
The first two terms in Eq. (\ref{Stres}) is the variation of the
action (\ref {Actg}) with respect to $\gamma _{ab}$, and the last
term is the counterterm which removes the divergences. The
conserved quantities associated to a Killing vector $\xi ^{a}$ is
\begin{equation}
{\cal Q}(\xi )=\int_{{\cal B}}d^{n}x\sqrt{\sigma }n^{a}T_{ab}\xi
^{b},
\end{equation}
where $\sigma $ is the determinant of the metric $\sigma _{ij}$,
appearing in the ADM-like decomposition of the boundary metric
\begin{equation}
ds^{2}=-N^{2}dt^{2}+\sigma _{ab}(dx^{a}+N^{a}dt)(dx^{b}+N^{b}dt).
\end{equation}
For boundaries with timelike Killing vector ($\xi =\partial
/\partial t$) and rotational Killing vector field $(\zeta
=\partial /\partial \phi )$ one obtains the conserved mass and
angular momentum of the system enclosed by the boundary ${\cal
B}$. In the context of AdS/CFT correspondence, the limit in which
the boundary ${\cal B}$ becomes infinite $({\cal B}_{\infty })$ is
taken, and the counterterm prescription ensures that the action
and conserved charges are finite. Using the above definition for
the conserved
charges, we find the total energy and angular momenta per unit volume $%
V_{n-1}$ of the solution as
\begin{eqnarray}
M &=&\frac{1}{16\pi }m\left[ n\Xi ^{2}-1\right] ,  \label{Mass} \\
J_{i} &=&\frac{1}{16\pi }n\Xi ma_{i}.  \label{Angmom}
\end{eqnarray}

\section{Thermodynamics of the Black Branes}

\label{Therm}

Calculating all the thermodynamic and conserved quantities of the
black brane solutions, we now check the first law of
thermodynamics for our solutions. We first obtain the mass as a
function of the extensive quantities $S$, ${\bf J}$, and $Q$.
Using the expression for the entropy, the mass, the angular
momenta, and the charge given in Eqs. (\ref{Entropy}),
(\ref{Charg}), (\ref{Mass}), (\ref{Angmom}), and the fact that
$f(r_{+})=0$, one can obtain a Smarr-type formula as
\begin{equation}
M(S,{\bf J},Q)=\frac{(nZ-1)J}{nl\sqrt{Z(Z-1)}},  \label{Smar}
\end{equation}
where $J=\left| {\bf J}\right| =\sqrt{\sum_{i}^{k}J_{i}^{2}}$ and
$Z=\Xi ^{2} $ is the positive real root of the following equation:

\begin{eqnarray}
&&\left( n(n-1)+4\beta ^{2}l^{2}\right) S^{n/(n-1)}-4\beta
l^{2}S^{1/(n-1)}\sqrt{\pi ^{2}Q^{2}+\beta
^{2}S^{2}}-\frac{4^{(n-2)/(n-1)}(n-1)\pi
lJZ^{1/(2n-2)}}{\sqrt{(Z-1)}}  \nonumber \\
&&+4\frac{n-1}{n-2}\pi ^{2}l^{2}Q^{2}S^{-(n-2)/(n-1)}\ {_{2}F_{1}\left( %
\left[\frac{1}{2},\frac{n-2}{2n-2}\right],\left[\frac{3n-4}{2n-2}\right],-{%
\frac{\pi ^{2}Q^{2}}{\beta ^{2}S^{2}}}\right)} =0.  \label{Zsmar}
\end{eqnarray}
One may then regard the parameters $S$, $J_{i}$'s, and $Q$ as a
complete set of extensive parameters for the mass $M(S,{\bf J},Q)$
and define the intensive parameters conjugate to them. These
quantities are the temperature, the angular velocities, and the
electric potential
\begin{equation}
T=\left( \frac{\partial M}{\partial S}\right) _{J,Q},\ \ \Omega
_{i}=\left(
\frac{\partial M}{\partial J_{i}}\right) _{S,Q},\ \ \Phi =\left( \frac{%
\partial M}{\partial Q}\right) _{S,J}.  \label{Dsmar}
\end{equation}
It is a matter of straightforward calculation to show that the
intensive
quantities calculated by Eq. (\ref{Dsmar}) coincide with Eqs. (\ref{Temp}), (%
\ref{Om}), and (\ref{Pot}). Thus, these quantities satisfy the
first law of thermodynamics
\[
dM=TdS+{{{\sum_{i=1}^{k}}}}\Omega _{i}dJ_{i}+\Phi dQ.
\]

\subsection{Stability in the canonical and the grand-canonical ensemble}

\begin{figure}[tbp]
\epsfxsize=10cm \centerline{\epsffile{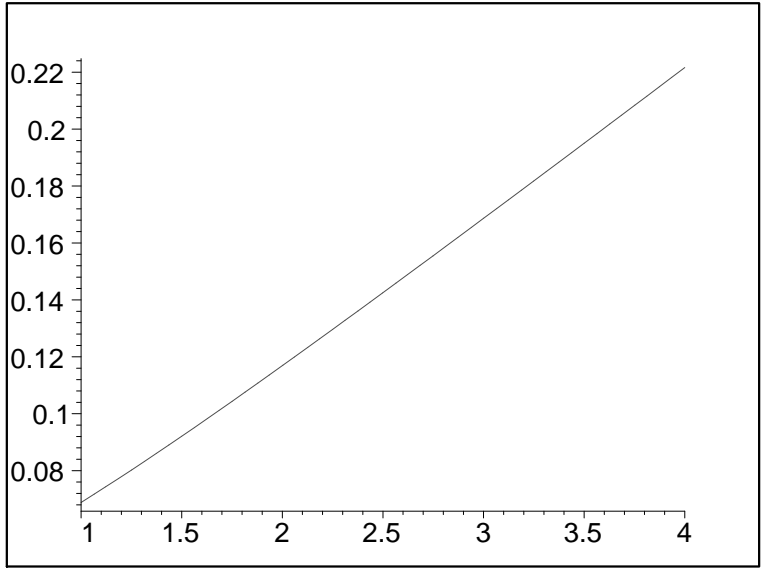}}\caption{$C_{{\bf
J},Q}$ versus $\Xi $ for $n=4 $, $l=1$, $\protect\beta =1$,
$q=0.7$ and $r=0.8$.} \label{Fig1}
\end{figure}

Next, we investigate the stability of charged rotating black brane
solutions of Born-Infeld gravity. The stability of a thermodynamic
system with respect to small variations of the thermodynamic
coordinates is usually performed by analyzing the behavior of the
entropy $S(M,{\bf J},Q)$ around the equilibrium. The stability can
also be studied by the behavior of the energy $M(S,{\bf J},Q)$
which should be a convex function of its extensive variable. Thus,
the local stability can in principle be carried out by finding the
determinant of the Hessian matrix of $M(S,Q,{\bf J})$ with
respect to its extensive variables $X_{i}$, ${\bf H}_{X_{i}X_{j}}^{M}=[%
\partial ^{2}M/\partial X_{i}\partial X_{j}]$ \cite{Gub}. In our case the
mass $M$ is a function of entropy, angular momenta, and charge.
The number of thermodynamic variables depends on the ensemble that
is used. In the canonical ensemble, the charge and the angular
momenta are fixed parameters, and therefore the positivity of the
heat capacity $C_{{\bf J},Q}=T_+/(\partial ^{2}M/\partial S^{2})_{{\bf J}%
,Q}$ is sufficient to ensure local stability. $(\partial
^{2}M/\partial S^{2})_{{\bf J},Q}$ at constant charge and angular
momenta is
\begin{eqnarray}
\left( \frac{\partial ^{2}M}{\partial S^{2}}\right) _{_{{\bf J},Q}} &=&\Upsilon^{-1}\Bigg\{\frac{%
\left( n-1\right) l^{2}m}{ r_{+}^{n}}\left[ (n-2)\Xi ^{2}+1\right]
\times  \Bigg. \nonumber \\
&&\left\{ n(n-1)\sqrt{1+\eta _{+}}+4\left( n-2\right) l^{2}{\beta
}^{2}\eta_+
+4l^{2}\beta ^{2}\left( \sqrt{1+\eta _{+}}-1\right) \right\} + \Bigg. \nonumber \\
&&n\Big\{32l^{4}\beta ^{4}\left( \sqrt{1+\eta _{+}}-1\right)
+n\left( n-1\right) \left( n(n-1) \sqrt{1+\eta _{+}}-8l^{2}{%
\beta }^{2}\eta _{+}\right) +  \nonumber \Big. \Bigg.\\
&&8nl^{2}\beta ^{2}\left[ 2l^{2}{\beta }^{2}\eta _{+}\left( \sqrt{1+\eta _{+}%
}-2\right) +n(n-1)\left( \sqrt{1+\eta _{+}}-1\right) \right]
\Big\}(\Xi ^{2}-1)\Bigg\}; \nonumber\\
\Upsilon &=&(n-1)^{3}\pi m\Xi ^{2}l^{4}\left[ (n-2)\Xi
^{2}+1\right] \left( 1+\eta _{+}\right) ^{1/2}r_{+}^{2}
\label{dMSS}
\end{eqnarray}
The heat capacity is positive for $m\geq m_{\rm ext}$, where the
temperature is positive. This fact can be seen easily for $\Xi
=1$, where the second term of Eq. (\ref{dMSS}) is zero and the
first term is positive. Also, one may see from Fig. \ref{Fig1}
that the heat capacity increases as $\Xi$ increases, and therefore
it is always positive. Thus, the black brane is stable in the
canonical ensemble. In the grand-canonical ensemble, after some
algebraic manipulation, we obtain
\begin{equation}
H_{S{\bf J}Q}^{M}=\frac{\Pi }{\Psi }
\end{equation}
where
\begin{eqnarray}
\Pi  &=&\frac{64\pi }{l^{2}\Xi ^{6}[(n-2)\Xi ^{2}+1]}\Bigg\{\left(
2\beta ^{2}(\left( 1+\eta _{+}\right)
^{1/2}-1)+(n-1)(n-2)^{2}q^{2}r_{+}^{-2n+2}{-\Lambda }\left( 1+\eta
_{+}\right) ^{1/2}\right)\Bigg.   \nonumber \\
&&\times {_{2}F_{1}\left( \left[ \frac{1}{2},\frac{n-2}{2n-2}\right] ,\left[ \frac{%
3n-4}{2n-2}\right] ,-\eta _{+}\right)} +(n-2)\newline \left(
(2\beta ^{2}\newline -\Lambda ){-2}\beta ^{2}\left( 1+\eta
_{+}\right) ^{1/2}\right) \Bigg\}
\end{eqnarray}
and
\begin{eqnarray}
\Psi  &=&(n-1)(n-2) r_{+}^{n-2}\left( 1+\eta _{+}\right)
^{1/2}\Bigg\{(n-1)^{2}q^{2}
{_{2}F_{1}\left( \left[ \frac{1}{2},\frac{n-2}{2n-2}\right] ,\left[ \frac{3n-4%
}{2n-2}\right] ,-\eta _{+}\right)} \Bigg.  \nonumber \\
&&+r_{+}^{2n-2}\left( (2\beta ^{2}-\Lambda )-2\beta ^{2}\left(
1+\eta _{+}\right) ^{1/2}\right) \Bigg\},
\end{eqnarray}
Since the value of $_{2}F_{1}\left( \left[ \frac{1}{2},\frac{n-2}{2n-2}%
\right] ,\left[ \frac{3n-4}{2n-2}\right] ,-\eta_+ \right) $ is
between $0$ and $1$ and $2\beta ^{2}\left( 1+\eta _{+}\right)
^{1/2}\leq (2\beta
^{2}-\Lambda )$ for $q\leq q_{{\rm ext}}$, it is easy to see that $H_{S{\bf J%
}Q}^{M}$\ is positive for all the allowed values of
\begin{equation}
q\leq q_{{\rm ext}}=\frac{r_{+}^{n-1}}{\beta }\sqrt{\frac{\Lambda
(\Lambda -4\beta ^{2})}{2(n-1)(n-2)}}  \label{qless}
\end{equation}
Thus, the $(n+1)$-dimensional AAdS charged rotating black brane is
locally stable in the grand-canonical ensemble.
\section{CLOSING REMARKS}

In this paper, we introduced a new class of charged, rotating solutions of $%
(n+1)$-dimensional Einstein-Born-Infeld gravity and investigate
their properties. These solutions are asymptotically anti-de
Sitter. In the
particular case $\beta \rightarrow \infty $, these solutions reduce to the $%
(n+1)$-dimensional charged rotating black brane solutions given in
Ref. \cite {Deh2}. We found that these solutions can represent
black brane, with inner and outer event horizons, an extreme black
brane or a naked singularity provided the parameters of the
solutions are chosen suitably. We also computed thermodynamic
quantities of the $(n+1)$-dimensional rotating charged black brane
such as the temperature, entropy, charge, electric potential, mass
and angular momentum and found that they satisfy the first law of
thermodynamics. We found that these thermodynamic quantities are
independent of the Born-Infeld parameter $\beta $.

Also, we studied the phase behavior of the charged rotating black branes in $%
(n+1)$ dimensions and showed that there is no Hawking-Page phase
transition in spite of the charge and angular momenta of the
branes. Indeed, we
calculated the heat capacity and the determinant of the Hessian matrix of $%
M(S,{\bf J},Q)$ with respect to its extensive variables $S$, ${\bf J}$ and $%
Q $ of the black brane and found that they are positive for all
the phase space, which means that the brane is stable for all the
allowed values of the metric parameters. This phase behavior is in
commensurable with the fact that there is no Hawking-Page
transition for black object whose horizon is diffeomorphic to
${\Bbb R}^{p}$ and therefore the system is always in the high
temperature phase \cite{Wit}.

\acknowledgements{This work has been supported by Research
Institute for Astrophysics and Astronomy of Maragha, Iran.}

\end{document}